\def\beg{\begin{equation}}
\def\eeq{\end{equation}}
\documentstyle[12pt]{article}
\begin{document}
\begin{center}
{\Large{\bf Thermoelectric power in between two plateaus in quantum 
Hall effect.}}
\vskip0.35cm
{\bf Keshav N. Shrivastava}
\vskip0.25cm
{\it School of Physics, University of Hyderabad,\\
Hyderabad  500046, India}
\end{center}

We have considered the response at two energies corresponding to two 
plateaus in the quantum Hall effect. Since the thermoelectric power
 involves the derivative of conductivity with respect to energy, we 
introduce the concept of a line width and hence an activation energy. 
We then use the Hall conductivity to define the thermoelectric power
 at the centre of two plateaus, which is found to vary monotonically 
as T$^2$ at low temperatures with fixed magnetic field. At elivated 
temperatures, the thermoelectric power varies as T exp(-$\Delta/k_B$T).
\vskip0.1cm
\noindent({\it Key words:} Thermoelectric power, Hall effect, Mott's 
formula.)\\
Corresponding author: keshav@mailaps.org\\
Fax: +91-40-2301 0145.Phone: 2301 0811.
\vskip1.0cm

\noindent {\bf 1.~ Introduction}

     The phenomenon of quantum Hall effect has been described for a 
long time[1,2]. The thermoelectric power involves the derivative of 
conductivity as a function of energy so that at the plateau in the 
Hall resistivity, where energy is fixed, it is not well defined. On 
the
 other hand, if we look at the centre of two plateaus, at a fixed 
magnetic field, we can define this derivative by using a relaxation 
process. The thermoelectric power can then give information about the
 relaxation mechanism in the quantum Hall effect away from the plateaus.
 In the transverse resistivity there are minima at the same magnetic 
fields where there are plateaus in the longitudinal resistivity. These
 fields are described by integers as well as fractions which are
 predicted by us[3].

     In this communication, we obtain the thermoelectric power at the 
centre of two plateaus. This is an important point because the system
 becomes metallic just before the plateaus and hence there is a peak 
in the resistivity at a point in the centre of two plateaus. The system
 is some times referred to as ``insulator" near the peak. Thus we are 
able to find the thermoelectric power in the insulating phase.
 
\noindent{\bf 2.~~Theory}

We consider a three level system in which the energies are labeled as
 E$_o$, E$_1$ and E$_2$. It absorbs energies E$_2$-E$_o$ and E$_1$-$E_o$ 
so that there are two lines in the absorption as a function of energy. 
If these lines are far apart, there is no overlap and hence there is no 
region where the absorption depends on line width. We select a problem
 where there is a region in which the absorption depends on the line
 width. The width of the line at E$_1$ is $\delta\Omega_1$ and that at
 E$_2$ is $\delta\Omega_2$. The response function in this case is given 
by,
\beg
\chi^"={\delta\Omega_1\over \{(E-E_1)/\hbar\}^2+(\delta\Omega_1)^2}+
{\delta\Omega_2\over\{(E-E_2)/\hbar\}^2+(\delta\Omega_2)^2}.
\eeq
If $E_2-E_1>>\hbar\delta\Omega_1$, the lines are well resolved and
 peaked at well defined energies such as E$_1$ and E$_2$. On the other
 hand if $E_2-E_1< \hbar \delta\Omega_1$, the lines overlap and there 
is absorption in the centre of two spectral lines. Since we are 
interested in the thermoelectric power in the centre of two lines, 
we consider the case of overlap between two lines. We suppose that
 the two lines are separated by $\delta_1$ such that $\delta_1$ is
 smaller than either $\Omega_1$ or $\Omega_2$, i.e.,
\beg
E_2-E_1 = \delta_1
\eeq
\beg
\delta_1<< \delta\Omega_1
\eeq
\beg
\delta_1<< \delta\Omega_2.
\eeq
At $E=E_1$, the first term in the response function is peaked so 
that the response becomes,
\beg
\chi^" = {1\over \delta\Omega_1}+ { \delta\Omega_2\over
 \{(E_1-E_2)/\hbar\}^2+(\delta\Omega_2)^2}
\eeq
but the line at $E_2$ is very near the one at $E_1$ so that the lines
 overlap. Substituting (2) in (5) we find,
\beg
\chi^"={1\over \delta\Omega_1}+{\delta\Omega_2\over(\delta_1/\hbar)^2+
(\delta
\Omega_2)^2}.
\eeq
For convenience of algebra, we assume $\delta\Omega_1\simeq
 \delta\Omega_2$ so that the centre of the lines between $E_1$ and
 $E_2$ is located at $E_1+{1\over 2}\delta_1$. If the relaxation 
occurs due to interaction with phonons, the width depends on the 
phonon correlation function, $2n+1$, where
 $n=[exp(\Delta/k_BT)-1]^{-1}$. It is also possible to say that
 $n\simeq exp(-\Delta/k_BT)$ at 
low temperatures so that $\Delta$ becomes the activation energy,
\beg
\delta\Omega_1 = c_1(2n+1) \simeq 2c_1exp(-\Delta/k_BT)
\eeq
In the symmetric case, 
\beg
\delta\Omega_1 = \delta_1.
\eeq
Let us consider the old Hall effect. In the quantum Hall effect 
the conductivity is quantized as $\sigma=ie^2/h$ where $i$ may be an
 integer or a fraction. We believe that this fraction depends on
 spin. If there are two plateaus, they occur at $i=\nu_1$ and
 $i=\nu_2$. The conductivity is quantized at both $\nu_1$ as well as
 at $\nu_2$ but we are neither interested in quantization at $\nu_1$ 
nor at $\nu_2$. We 
are
interested in the point which is exactly in the middle of $\nu_1$ and
 $\nu_2$ so that there is no quantization at this point.

     The current is determined by $j=nev$ where $n$ is the 
concentration of electrons, $e$ the charge and $v$ the velocity. We 
replace the charge by energy $E_c-E_F+{3\over 2}k_BT$ and devide the
 resulting expression by the current to define the Peltier coefficient,
\beg
\Pi=- {E_c-E_F+{3\over 2}k_BT\over e}
\eeq
where $E_c$ is the electron energy in the conduction band. The
 negative sign in the above has come from the sign of the charge, -e.
 The thermoelectric power, S, may be defined as,
\beg
\Pi = ST
\eeq
where T is the temperature. If $|\vec E|$ is the electric field, 
another
definition of the thermoelectric power is obtained by,
\beg
|\vec E| = S grad T
\eeq
or
\beg
S={|\vec E|\over grad T}.
\eeq
We can replace the $grad$ by $d/dx$ but then the temperature gradient
 has 
to be evaluated. Apparently, Mott has solved[4] this problem by 
defining the thermoelectric power by using the energy derivative of 
the logarithm of the conductivity,
\beg
S={\pi^2k_B^2T\over 3e}{d\over dE} {\it l}n \sigma|_{E=E_F}.
\eeq
This means that we should know the conductivity as a function of 
energy, then only we can use the Mott formula. As we described above, 
we are interested in a point which is mid way between two quantized 
values, we use the old result according to which the Hall resistivity 
is $\rho=nec/B$ and hence the Hall conductivity is,
\beg
\sigma = {nec \over B}
\eeq
where $n$ is the electron concentration, $e$ is the charge, $B$ is 
the magnetic field and c is the velocity of light. The transition 
energies in the system described above are $E_1={1\over 2} g\mu_BB_1$ 
and $E_2={1\over 2}g\mu_BB_2$. Our transitions occur at $E_1$ and 
$E_2$ and the centre is at $E_1+{1\over 2}\delta_1$. Thus there is 
quantization at $E_1$ and at $E_2$ but not at $E_1+{1\over 2}\delta_1$.
Therefore, the conductivity becomes,
\beg
\sigma={1\over 2}g\mu_B nec {1\over E}
\eeq
where $E = {1\over 2}g\mu_BB$. Taking the logarithm of conductivity,
 we obtain,
\beg
{\it l}n \sigma={\it l}n({1\over 2}g\mu_Bnec)-{\it l}n E
\eeq
so that
\beg
{d\over dE}{\it l}n \sigma=-{d\over dE}{\it l}n E= - {1\over E}.
\eeq
If the conductivity is thermally activated,
\beg
\sigma = \sigma_o exp(E/k_BT)
\eeq
we obtain,
\beg
{d\over dE}{\it l}n\sigma={1\over k_BT}.
\eeq
In this case, the thermoelectric power will be independent of 
temperature. Therefore, we do not consider this case but we consider 
the thermally activated relaxation rate. We go back to the Mott 
formula to find the thermoelectric power as,
\beg
S=-{\pi^2k_B^2T\over 3e}{1\over E}.
\eeq
Where the negative sign means that power reduces as the energy is
 increased. When energy is either $E_1$ or $E_2$, the conductivity 
is quantized but at the mid point, $E=E_1+{1\over 2}\delta_1$, the 
conductivity is not quantized so that at this point,
\beg
S=-  {\pi^2k_B^2T\over 3eE_1(1+{1\over 2}{\delta_1\over E_1})}.
\eeq
If $\delta_1$ is very small compared with $E_1$ we can write the term 
in the denominator as,
\beg
S = {\delta_1\pi^2k_B^2T\over 6eE_1^2} -{\pi^2k_B^2T\over 3eE_1}.
\eeq
Thus there is a large negative term which is linear in T and a
 small correction which depends on the relaxation[5]. The distance
 ${1\over 2}\delta_1$ must be equal to half width so that,
\beg
\delta_1={1\over 2\tau_1}=c_1(2n+1).
\eeq
If the system is thermally activated, $2n+1\simeq 2n \simeq 
2exp(-\Delta/k_BT)$. Therefore, the first term in the thermoelectric
 power varies as,
\beg
S_1= {\pi^2k_B^2Tc_1(2n+1)\over 6eE_1^2}\simeq {\pi^2k_B^2c_1
Texp(-\Delta/k_BT)\over 3eE_1^2}
\eeq
in which the temperature dependence is determined by
\beg
S_1\simeq c_2Tcoth{\Delta\over 2k_BT}
\eeq
which at high temperatures varies as,
\beg
S_1\simeq {2c_2k_BT^2\over \Delta}.
\eeq
There is also a large negative term,
\beg
S= S_1 + S_2
\eeq
\beg
S_2 = - {\pi^2k_B^2T\over 3eE_1}.
\eeq

Thus we have two terms in the thermoelectric power. One of these
 terms is negative and linear in temperature. This describes the 
emission of power under favourable conditions such as large 
relaxation times in the excited states. The other term is positive 
and depends approximately on $T^2$. It may be described as
 $T coth(\Delta/2k_BT)$. It is known that relaxation in quantum Hall 
effect is thermally activated so that the thermopower may appear as 
$T exp(-\Delta/k_BT)$ as given by (24). In these results, the 
magnetic field is held constant so that there is no phase transition
 by varying temperature.

     The variation of conductivity as a function of length or 
thickness
of the sample is well known and it is related to a critical 
exponent.
Polyakov and Shklovskii[6] have treated $\delta_1$ or an equivalent 
quantity as if there is a phase transition with a critical
 temperature. In that case, the width is described by
 ``scaling exponents",
\beg
\delta_1 \simeq T^\kappa
\eeq
where $\kappa \simeq$ 0.4. It has been possible to correlate the
 exponent $\kappa$ with $\gamma$ which describes the coherence 
length. The band filling factor is called $\nu$ and the coherence
 length 
varies as,
\beg
\xi(\nu)=\xi(0)|\nu - \nu_o|^{-\gamma}
\eeq
with $\gamma$ = 2.3 or $\kappa=1/\gamma$=0.435. Long time ago[5],
 the value of $\kappa$ was thought to be 0.5. We write the scaling 
width as,
\beg
\delta_1=c_2T^\kappa
\eeq
which shows that the positive term of the thermoelectric power 
varies as,
\beg
S_1= {\pi^2k_B^2c_2\over 6eE_1^2}T^{\kappa +1}
\eeq
so that the exponent $\kappa$+1 can be determined from the 
thermoelectric power as a function of temperature. The critical
 exponents are a characteristic of a phase transition. When 
$T_c\simeq$0, $(T-T_c)^\kappa =T^\kappa$ so that Polyakov's result
 is applicable to a phase transition at zero temperature. When we 
go from one level to another, we call it a phase transition. However,
 if we keep the magnetic field a constant at the middle of two 
plateaus, then there is no phase transition and we need not expect
 any critical  exponent.

     Possanzini et al[7] have measured the thermoelectric power for
 two different densities. In both the cases, the thermoelectric power 
monotonically increases with increasing temperature from 0.4 to 4 K.
 An effort is made to fit the data with $S=\alpha T^{1/3}+\eta T^4$ 
which is interpreted to arise from a variable range hopping model or
 from $S_{xx}=\gamma +\eta T^4$. Naturally, a term can be added in 
which the exponent has any value between 1/3 and 4 without 
disturbing the agreement between the model and the data. The 
expression (26) shows that the thermoelectric power varies as $T^2$.
 When the magnetic field is kept constant and the temperature is 
varied, it is reasonable to expect that there is no phase transition
 so the thermoelectric power varies as $T^2$. At a temperature of
 0.32 K, the largest positive value of the diagonal thermoelectric
 power is about 5$\mu$V/K and at very small magnetic fields, the 
value is negative of the order of 10$\mu$V/K. Thus these observed 
features are in accord with the theoretical expressions. 
   
\noindent{\bf3.~~ Conclusions}.

     In conclusions,we find that when field is fixed in between 
two plateaus, the thermoelectric power varies as $T^2$ at low
 temperatures. At slightly elivated temperatures,we expect a
 thermally activated relaxation process so that
 $T exp(-\Delta/k_BT)$ type variation is predicted.

\vskip1.25cm

\noindent{\bf4.~~References}
\begin{enumerate}
\item S. Das Sarma and A. Pinczuk, eds., Perspectives in quantum 
Hall effect, Wiley, New York 1997.
\item R. E. Prange and S. M. Girvin, eds., The quantum Hall effect,
 Springer-Verlag, New York 1987.
\item K. N. Shrivastava, Introduction to quantum Hall effect, Nova 
Science, New York 2002.(See also cond-mat/0303309).
\item N. F. Mott, J. Non-Cryst. Solids, {\bf1}, 1 (1968).
\item K. N. Shrivastava, Phys. Stat. Solidi (b){\bf 117},437(1983)
\item D. G. Polyakov and B. I. Shklovskii, Phys. Rev. B
 {\bf 48}, 11167 (1993).
\item C. Possanzini, R. Fletcher, P. T. Coleridge, Y. Feng, R. L.
 Williams and J. C. Maan, Phys. Rev. Lett. {\bf 90}, 176601(2003).
\end{enumerate}
\vskip0.1cm

\end{document}